\documentclass[fleqn,usenatbib]{mnras}

\usepackage{newtxtext,newtxmath}

\usepackage[T1]{fontenc}
\usepackage{ae,aecompl}

\usepackage{graphicx}	
\usepackage{amsmath}	
\usepackage{amssymb}	
\usepackage{soul}
\usepackage{tablefootnote}
\usepackage{seqsplit}

\title[Outflows and inflows in GRS~1716-249]{Discovery of optical outflows and inflows in the black hole candidate GRS~1716-249}
\author[V. A. C\'uneo et al.]{
V. A. C\'uneo,$^{1,2}$\thanks{E-mail: virginiacuneo@gmail.com}
T. Mu\~noz-Darias,$^{1,2}$ 
J. S\'anchez-Sierras,$^{1,2}$ 
F. Jim\'enez-Ibarra,$^{1,2}$ 
\newauthor 
M. Armas Padilla,$^{1,2}$ 
D. A. H. Buckley,$^{3}$ 
J. Casares,$^{1,2}$ 
P. Charles,$^{4}$ 
\newauthor 
J. M. Corral-Santana,$^{5,6}$ 
R. Fender,$^{7}$ 
J. A. Fern\'andez-Ontiveros,$^{8}$ 
D. Mata S\'anchez,$^{9}$ 
\newauthor 
G. Panizo-Espinar,$^{1,2}$
G. Ponti,$^{10,11}$ and 
M. A. P. Torres$^{1,2}$
\newauthor 
\\
$^{1}$Instituto de Astrof\'isica de Canarias (IAC), E-38205 La Laguna, Tenerife, Spain \\
$^{2}$Departamento de Astrof\'isica, Universidad de La Laguna, E-38206 La Laguna, Tenerife, Spain \\
$^{3}$South African Astronomical Observatory, PO Box 9, Observatory 7935, Cape Town, South Africa \\
$^{4}$Department of Physics \& Astronomy, University of Southampton, Southampton SO17 1BJ, UK \\
$^{5}$European Southern Observatory (ESO), Alonso de C\'ordova 3107, Vitacura, Casilla 19, Santiago, Chile \\
$^{6}$Pontificia Universidad Cat\'olica de Chile, Vicu\~na-Mackenna 4860, Macul, Santiago, Chile \\
$^{7}$Astrophysics, Department of Physics, University of Oxford, Keble Road, Oxford OX1 3RH, UK \\
$^{8}$Istituto di Astrofisica e Planetologia Spaziali (INAF-IAPS), Via Fosso del Cavaliere 100, 00133 Roma, Italy \\
$^{9}$Jodrell Bank Centre for Astrophysics, Department of Physics and Astronomy, The University of Manchester, Manchester M13 9PL, UK \\
$^{10}$INAF-Osservatorio Astronomico di Brera, Via E. Bianchi 46, I-23807 Merate (LC), Italy \\
$^{11}$Max-Planck-Institut f\"ur Extraterrestrische Physik, Giessenbachstrasse, D-85748, Garching, Germany \\
}

\date{Accepted XXX. Received YYY; in original form ZZZ}

\pubyear{2020}

\begin{document}
\label{firstpage}
\pagerange{\pageref{firstpage}--\pageref{lastpage}}
\maketitle

\begin{abstract}
We present optical spectroscopy obtained with the GTC, VLT and SALT telescopes during the decline of the 2016--2017 outburst of the black hole candidate GRS~1716-249 (Nova Oph 1993). Our 18-epoch data set spans 6 months and reveals that the observational properties of the main emission lines are very variable, even on time scales of a few hours. Several epochs are characterized by P-Cyg (as well as flat-top and asymmetric) profiles in the H$\alpha$, H$\beta$ and \ion{He}{ii} ($\lambda$4686) emission lines, implying the presence of an accretion disc wind, which is likely hot and dense. The wind's terminal velocity ($\sim$2000 km s$^{-1}$) is similar to that observed in other black hole X-ray transients. These lines also show transient and sharp red-shifted absorptions, taking the form of inverted P-Cyg profiles. We argue that these profiles can be explained by the presence of infalling material at $\sim$1300 km s$^{-1}$. We propose a failed wind scenario to explain this inflow and discuss other alternatives, such as obscuration produced by an accretion-related structure (e.g. the gas stream) in a high inclination system.
\end{abstract}




\section{Introduction}
Low mass X-ray binaries consist of a compact object, a neutron star or a black hole (BH), that accretes matter from a companion star via an accretion disc. A subclass known as X-ray binary transients (XRTs) spend most of their lives in a quiescent state, characterized by X-ray luminosities of $\sim 10^{31-34}$ erg s$^{-1}$. Eventually, these systems undergo outburst episodes, caused by an increase of the mass accretion rate, reaching X-ray luminosities of $\sim 10^{36-39}$ erg s$^{-1}$. XRTs exhibit different X-ray states during outburst \citep[e.g.][]{2006csxs.book..157M,2011BASI...39..409B}. In the hard state, Comptonized emission from a hot electron corona dominates the X-ray spectrum \citep[e.g.][]{1980A&A....86..121S,2007A&ARv..15....1D,2010LNP...794...17G}, while the soft state is dominated by thermal emission from the accretion disc \citep{1973A&A....24..337S}. The hardness-intensity diagram is arguably the main tool to determine the evolution of a  source during outburst \citep{2001ApJS..132..377H}. XRTs typically evolve from the hard state to the soft state and back to the hard state, describing a ``q" shape in this diagram, with the hard-to-soft transition occurring at higher luminosity than the soft-to-hard \citep[e.g.][]{2005A&A...440..207B}.

Different types of outflows are associated with XRTs during outburst. Hot, X-ray winds of mid to highly ionized material are typically observed in the soft state of high inclination systems \citep{2006Natur.441..953M,2009Natur.458..481N,2012MNRAS.422L..11P,2016AN....337..512P,2016AN....337..368D}. On the other hand, the hard state is characterized by the presence of collimated radio jets \citep{1999ARA&A..37..409M,2004MNRAS.355.1105F}. In addition, the BH transient V404 Cyg exhibited P-Cyg profiles in a number of H and He optical lines during its 2015 outburst, implying the presence of a low-ionisation accretion disc wind \citep{2016Natur.534...75M}. This outflow, unusually for X-ray winds, was observed simultaneously with the radio jet. Optical accretion disc winds have been detected in a number of XRTs since then \citep[e.g.][]{2019ApJ...879L...4M}, mainly in the H$\alpha$ and \ion{He}{i} ($\lambda$5876 and $\lambda$6678) emission lines. In addition,  broad emission line wings and absorption troughs, superimposed on the standard double peak profile produced by the accretion disc rotation \citep{1969AcA....19..155S}, have been found in some systems \citep[e.g.][]{2019ApJ...879L...4M}. These features are often observed simultaneously with the P-Cyg profiles \citep[e.g. fig. 15 in][]{2018MNRAS.481.2646M} and arise most likely in the wind \citep[see e.g.][for a discussion]{2020arXiv200707257S}.

The XRT GRS~1716-249 (Nova Oph 1993) was discovered in September 1993 by BATSE/CGRO \citep{1993IAUC.5874....1B-Harmon} and SIGMA/GRANAT \citep{1993IAUC.5874....1B-Ballet}. After its discovery outburst, the source remained in quiescence for 23 years until a new outburst was detected by the Monitor of All-sky X-ray Image (\textit{MAXI}; \citealt{2009PASJ...61..999M}), on board the International Space Station, in December 16, 2016 \citep{2016ATel.9876....1N,2016ATel.9895....1M}. \citet{1996A&A...314..123M} determined a mass for the compact object of $>$~4.9~M$_{\odot}$, implying a BH nature, but this estimation is based on a tentative superhump period of $\sim$15 hours that requires confirmation. Nevertheless, some X-ray properties of the source, such as the horizontal transition from the hard to the soft state in the hardness-intensity diagram \citep[see fig. 2 in][]{2019MNRAS.482.1587B} and the detection of type-C quasi-periodic oscillations typical of BH XRTs \citep{1996ApJ...458L..75V,2019MNRAS.487.3150B}, also point to a BH accretor. 

In this work we analyse optical spectroscopy obtained during the 2016--2017 outburst of GRS~1716-249 to investigate the presence of accretion disc wind signatures in the main optical emission lines. We describe the observations in Section \ref{obs}. In Section \ref{a&r} we detail the analysis and results. We discuss our results in Section \ref{discussion} and summarise our main conclusions in Section \ref{summary}.

\section{Observations}
\label{obs}
We obtained optical spectroscopy in 18 epochs during 2017 using different telescopes and covering the decline of the outburst. The observing log is detailed in Table \ref{log}.

\begin{table*}
\caption{Observing log of GRS~1716-249 in 2017. }
\begin{tabular}{ccccccccc}
\hline 
Epoch & Date (hh:mm:ss) & MJD & Days since & Telescope & H$\alpha$ & H$\beta$ & \ion{He}{ii} & Magnitude \\
 &  &  & MJD 57680 &  &  &  & ($\lambda$4686) & \\
\hline
\rule{0pt}{3ex}1 & 27/02 (06:51:22) & 57811.286 & 131.286 & GTC &  & -- & -- & 17.49$\pm$0.01 (g) \\
\rule{0pt}{3ex}2 & 01/03 (06:33:09) & 57813.273 & 133.273 & GTC & ? &  &  & 16.799$\pm$0.005 (r) \\
\rule{0pt}{3ex}3 & 30/03 (00:41:30) & 57842.029 & 162.029 & SALT & P-Cyg$+$flat-top &  &  &  \\
\rule{0pt}{3ex}4 & 02/04 (04:59:37) & 57845.208 & 165.208 & GTC & iP-Cyg & iP-Cyg &  & 16.662$\pm$0.004 (r) \\
\rule{0pt}{3ex}5 & 03/04 (05:24:15) & 57846.225 & 166.225 & GTC & flat-top &  &  & 16.84$\pm$0.01 (r) \\
\rule{0pt}{3ex}6 & 05/04 (05:36:02) & 57848.233 & 168.233 & GTC &  &  &  & 16.851$\pm$0.005 (r) \\
\rule{0pt}{3ex}7 & 17/05 (01:45:03) & 57890.073 & 210.073 & GTC &  & -- & -- & 16.678$\pm$0.009 (r) \\
\rule{0pt}{3ex}8 & 18/05 (01:39:59) & 57891.069 & 211.069 & GTC &  & -- & -- & 16.94$\pm$0.02 (r) \\
\rule{0pt}{3ex}9 & 19/05 (03:05:42) & 57892.129 & 212.129 & VLT & iP-Cyg & iP-Cyg &  &  \\
\rule{0pt}{3ex}10 & 19/05 (21:19:22) & 57892.888 & 212.888 & SALT &  &  &  &  \\
\rule{0pt}{3ex}11 & 27/05 (03:32:32) & 57900.148 & 220.148 & GTC &  &  &  & 16.917$\pm$0.004 (r) \\
\rule{0pt}{3ex}12 & 03/06 (02:01:40) & 57907.084 & 227.084 & SALT &  & -- &  &  \\
\rule{0pt}{3ex}13-a & 03/06 (01:33:53) & 57907.065 & 227.065 & VLT & P-Cyg &  & P-Cyg* & 17.657$\pm$0.007 (g) \\
\rule{0pt}{3ex}13-b & 03/06 (02:06:27) & 57907.088 & 227.088 & VLT & P-Cyg &  & P-Cyg* &  \\
\rule{0pt}{3ex}13-c & 03/06 (02:42:41) & 57907.113 & 227.113 & VLT & P-Cyg$+$iP-Cyg & iP-Cyg & P-Cyg* &  \\
\rule{0pt}{3ex}13-d & 03/06 (03:15:15) & 57907.136 & 227.136 & VLT & P-Cyg$+$iP-Cyg & iP-Cyg & P-Cyg* &  \\
\rule{0pt}{3ex}13-e & 03/06 (03:51:32) & 57907.161 & 227.161 & VLT & P-Cyg &  & P-Cyg* &  \\
\rule{0pt}{3ex}13-f & 03/06 (04:24:05) & 57907.183 & 227.183 & VLT & P-Cyg &  & P-Cyg* &  \\
\rule{0pt}{3ex}13-g & 03/06 (05:38:09) & 57907.235 & 227.235 & VLT & P-Cyg & P-Cyg & P-Cyg* &  \\
\rule{0pt}{3ex}13-h & 03/06 (06:10:42) & 57907.257 & 227.257 & VLT & P-Cyg & P-Cyg & P-Cyg* &  \\
\rule{0pt}{3ex}13-i & 03/06 (06:47:25) & 57907.283 & 227.283 & VLT & P-Cyg & P-Cyg & P-Cyg* &  \\
\rule{0pt}{3ex}13-j & 03/06 (07:20:00) & 57907.306 & 227.306 & VLT & P-Cyg &  & P-Cyg* &  \\
\rule{0pt}{3ex}14 & 04/06 (03:03:14) & 57908.127 & 228.127 & GTC &  & -- & -- & 17.012$\pm$0.004 (r) \\
\rule{0pt}{3ex}15 & 14/06 (19:38:06) & 57918.818 & 238.818 & SALT &  &  & -- &  \\
\rule{0pt}{3ex}16 & 15/07 (23:17:18) & 57949.970 & 269.970 & GTC & ? & -- & -- & 17.543$\pm$0.007 (r) \\
\rule{0pt}{3ex}17 & 23/07 (22:36:43) & 57957.942 & 277.942 & SALT &  & -- & -- &  \\
\rule{0pt}{3ex}18 & 18/08 (21:04:40) & 57983.878 & 303.878 & GTC &  & -- & -- & 18.227$\pm$0.005 (r) \\
\hline
\multicolumn{9}{p{\dimexpr\textwidth-2\tabcolsep\relax}}{\seqsplit{\footnotesize{NOTE: The detection of P-Cyg, inverted P-Cyg (iP-Cyg) and flat-top profiles is indicated for each line. \ion{He}{ii} ($\lambda$4686) blue-shifted absorption components superimposed on the Bowen blend are marked as P-Cyg* (see Section \ref{winds}).}}} \\
\multicolumn{9}{l}{\footnotesize{? $=$ marginally consistent with the presence of a P-Cyg feature.}} \\
\multicolumn{9}{l}{\footnotesize{-- $=$ non detection of the line due to a poor signal-to-noise ratio.}}
\end{tabular}
\label{log} 
\end{table*}

\subsection{The Southern African Large Telescope (SALT)}
GRS~1716-249 was observed 5 nights by the Robert Stobie Spectrograph \citep{2003SPIE.4841.1463B}, attached to SALT \citep{2006SPIE.6267E..0ZB} at the South African Astronomical Observatory in Sutherland. We used the PG1300 and PG2300 gratings, with 1 and 1.5 arcsec slit-widths, which resulted in velocity resolutions of $\sim$300 km s$^{-1}$ and $\sim$230 km s$^{-1}$, respectively. A total of 39 spectra, with exposure times varying between 250 and 1200 s, were obtained. They were reduced using \textsc{IRAF}\footnote{IRAF is distributed by the National Optical Astronomy Observatory, which is operated by the Association of Universities for Research in Astronomy, Inc. under contract to the National Science Foundation.} and the \textsc{PySALT} package \citep{2016SPIE.9908E..2LC}.

\subsection{Gran Telescopio Canarias (GTC)}
We obtained 25 spectra on 11 different nights between February 27 and August 18, using the Optical System for Imaging and low-Intermediate-Resolution Integrated Spectroscopy \citep[OSIRIS;][]{2000SPIE.4008..623C}, attached to GTC at the Observatorio del Roque de los Muchachos in La Palma, Spain. The exposure times ranged between 200 and 1400 s. We used the R1000B (3630--7500 \AA), R2500V (4500--6000 \AA) and R2500R (5575--7685 \AA) grisms and slit-widths of 0.8--1~arcsec. These configurations resulted in velocity resolutions of 351--374 km s$^{-1}$, 116--154 km s$^{-1}$ and 111--196 km~s$^{-1}$, respectively. The reduction of the spectra was carried out with \textsc{IRAF}. 

\subsection{Very Large Telescope (VLT)}
The X-shooter spectrograph \citep{2011A&A...536A.105V}, attached to the VLT-UT2 at Cerro Paranal, Chile, observed GRS~1716-249 on the nights of May 19 and June 3. The spectrograph is equiped with 3 different arms to cover different spectral regions simultaneously. In this paper we use data from the ultraviolet and visible bands, which cover similar spectral ranges than the other facilities. We obtained a total of 22 spectra from each arm, with exposure times of 916--944~s, that were combined into 11 final spectra (one for the first night and 10 for the second one) on each spectral band. We used the 1 and 1.2 arcsec slits in the ultraviolet and visible arms, which provided velocity resolutions of 55 km s$^{-1}$ and 46 km s$^{-1}$, respectively. The spectra were processed using version 3.2.0 of the ESO X-shooter pipeline.  

\subsection{Photometry}
When available, we derived optical magnitudes from the acquisition images of each observing run. Flux calibration was performed against field stars catalogued in PanSTARRS \citep{2016arXiv161205242M} and using \textsc{astropy-photutils} based routines \citep{2019zndo...2533376B}. An averaged magnitude for each available night is reported in Table \ref{log}.

\section{Analysis and Results}
\label{a&r}
\begin{figure*}
 \includegraphics[trim=31mm 0mm 38mm 10mm,clip,width=\textwidth]{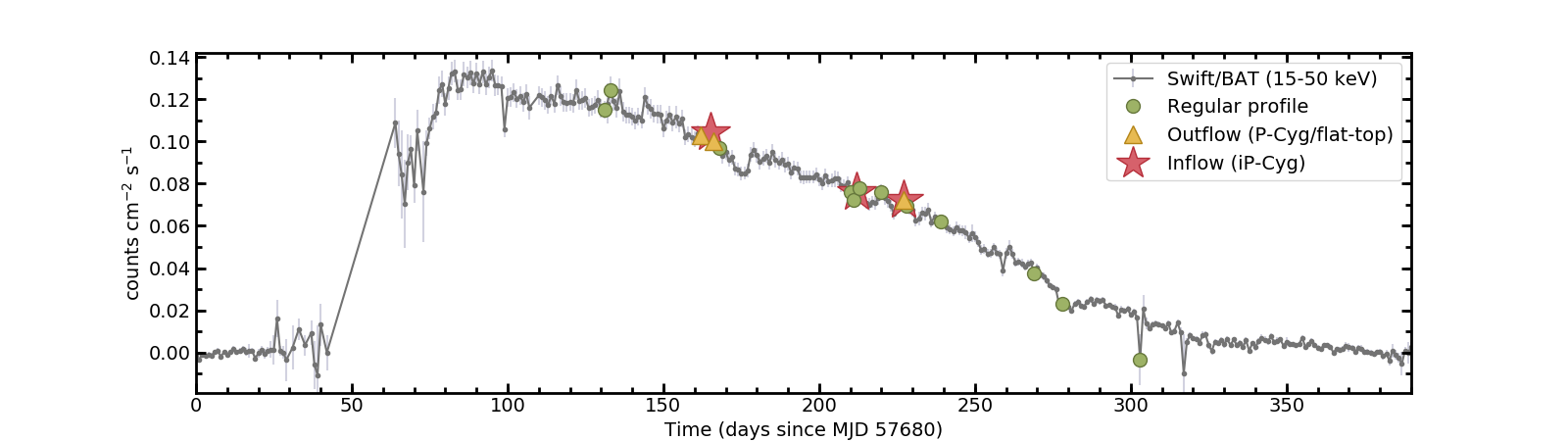}
 \caption{\textit{Swift}/BAT (15-50 keV) light curve of GRS~1716-249 during its 2016--2017 outburst. The epochs where outflow or inflow features were observed are marked by yellow triangles and red stars, respectively. Green dots correspond to spectroscopic epochs without outflow or inflow detections.}
 \label{lc}
\end{figure*}

In order to analyse the X-ray evolution of the source, we used daily averaged fluxes from the Burst Alert Telescope \citep[BAT;][]{2013ApJS..209...14K}, on board the Neil Gehrels \textit{Swift} Observatory \citep[\textit{Swift};][]{2004ApJ...611.1005G}, and from \textit{MAXI}. Fig. \ref{lc} shows the X-ray light curve in the 15-50 keV band, where we mark the epochs of our optical observations. All our spectra were obtained during the outburst decay phase.  

\begin{figure}
 \includegraphics[trim=4mm 5mm 10mm 10mm,width=\columnwidth]{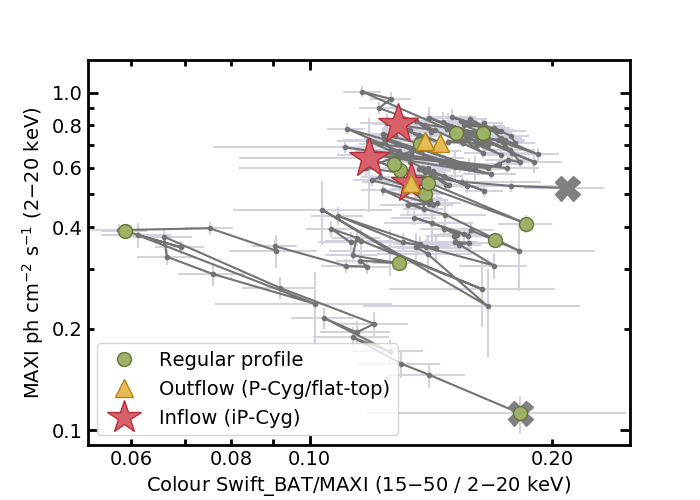}
 \caption{Hardness-intensity diagram of GRS~1716-249 during its 2016--2017 outburst from \textit{Swift}/BAT (15-50 keV) and \textit{MAXI} (2-20 keV) data. The colours and symbols are the same as in Fig. \ref{lc}. The top and bottom crosses denote the starting and last points, respectively.}
 \label{hid}
\end{figure}

Fig. \ref{hid} shows the hardness-intensity diagram. The X-ray colour is defined as the hard to soft ratio, with \textit{Swift}/BAT providing the hard (15--20 keV) band, and \textit{MAXI} the soft (2--20 keV) band. Only \textit{Swift}/BAT count rates higher than 0.02 counts cm$^{-2}$ s$^{-1}$ were considered. Most of the optical observations were taken during the hard state peak, but there is also one observation during a softening period and one in the low luminosity hard state.

We performed the analysis of the optical spectra using \textsc{molly} and custom software under \textsc{python}. In order to increase the signal-to-noise ratio (S/N), we nightly averaged the SALT and GTC spectra and rebinned the GTC (R2500V and R2500R) and all the SALT spectra, by factors of $\sim$2.6, $\sim$2 and 5, respectively. The range of values chosen is due to the different resolution of the spectra to be nightly averaged. We also rebinned the VLT spectra by a factor of 5, but due to the high variability present during epoch \#13 we analysed these spectra individually (see Table \ref{log}). We always achieved S/N values $>$ 20, with typical values $>$~70 and several epochs (including epoch \#13) above S/N $\sim$100.

The Balmer and \ion{He}{ii} transitions are observed in emission in all the spectra, while the main \ion{He}{i} transitions are too weak for a detailed study. Therefore, we focus our analysis on the Balmer series, particularly in H$\alpha$ and H$\beta$, and in the \ion{He}{ii} ($\lambda$4686) emission line. The spectral regions covering each line were normalized carefully and individually by fitting the adjacent continuum with a first order polynomial. H$\alpha$, H$\beta$ and \ion{He}{ii} ($\lambda$4686) emission lines exhibit strong variability throughout the observed period, including P-Cyg profiles, asymmetries, underlying broad absorptions and narrower red-shifted absorptions. The analysis of the line profile of \ion{He}{ii} ($\lambda$4686) is particularly complex due to the partial overlap with the Bowen blend (\ion{N}{iii}/\ion{C}{iii}/\ion{O}{ii}) at $\lambda$4640--4650. We describe these features in the following sections. For comparison with other XRTs, the H$\alpha$ profile has Full Width at Half Maximum (FWHM) values that increased slightly from $\sim$950 to $\sim$1400~km~s$^{-1}$ during our observing period, while the Equivalent Width (EW) remained approximately stable at a value of $\sim$4 \AA.

\subsection{P-Cyg profiles and other line asymmetries}
\label{winds}
\begin{figure}
    \centering
    \includegraphics[trim=5mm 48mm 10mm 57mm,width=0.99\columnwidth]{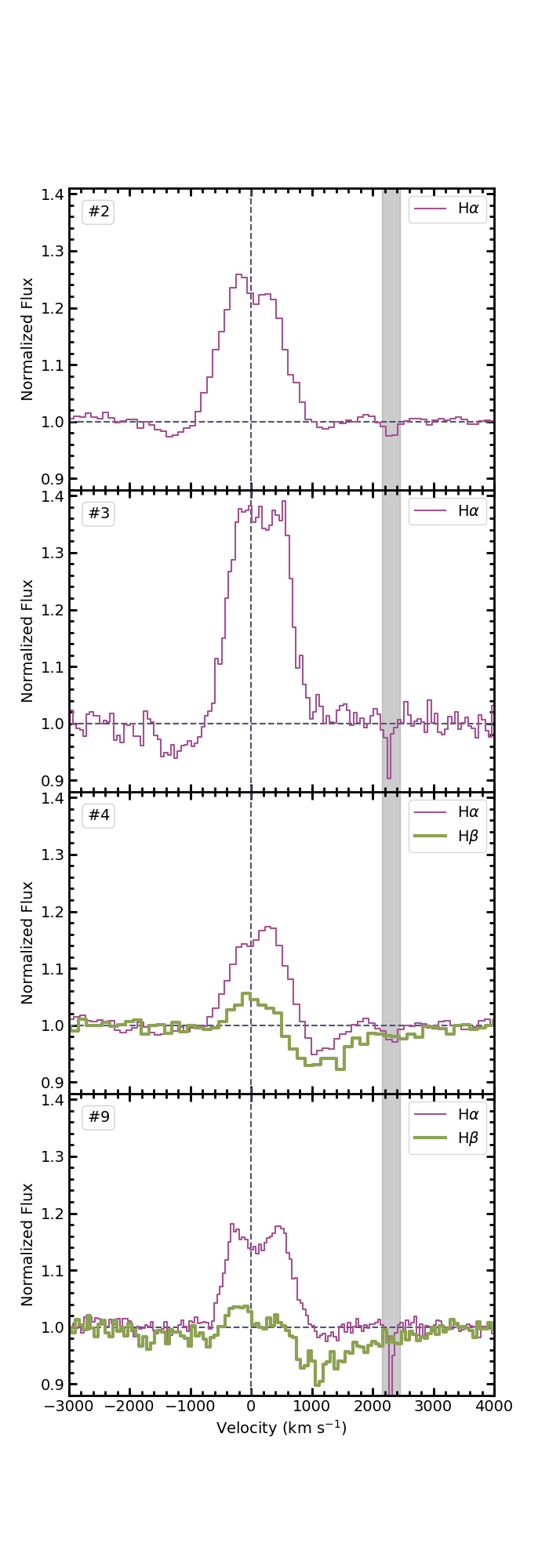}
    \caption{Examples of the H$\alpha$ (purple) and H$\beta$ (green) emission line profiles. A shallow P-Cyg profile can be seen at epoch \#2, a deeper one at epoch \#3 (together with a flat-top profile), and inverted P-Cyg profiles at epochs \#4 and \#9. The grey shaded band indicates contamination by interstellar features.}
    \label{sample}
\end{figure}

\begin{figure*}
 \includegraphics[trim=0mm 3mm 0mm 0mm,clip,width=\textwidth]{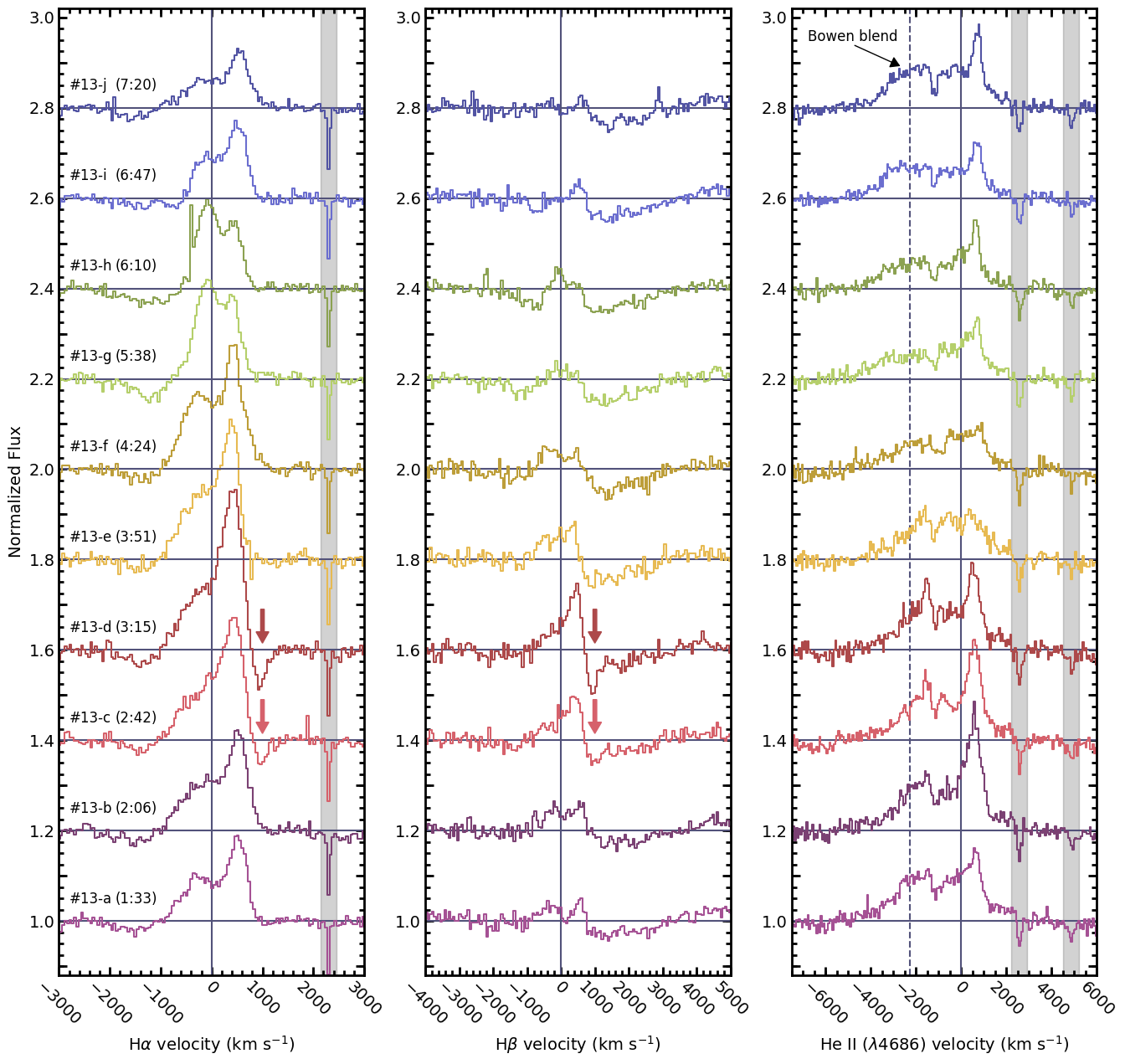}
 \caption{From bottom to top, evolution of the H$\alpha$ (left), H$\beta$ (centre) and \ion{He}{ii} ($\lambda$4686; right) emission line profiles, in the night of June 3 (epoch \#13); UTC is indicated in parenthesis. An offset of 0.2 in the y-axis was applied for clarity. The grey shaded bands indicate contamination by interstellar features. The arrows mark the inflow features. The dashed line denotes the \ion{C}{iii} ($\lambda$4650.1) line of the Bowen blend for reference.}
 \label{vlt}
\end{figure*}

P-Cyg profiles are features that have long been associated with wind-type outflows \citep{1929MNRAS..90..202B,1979ApJS...39..481C}. However, accretion discs in X-ray binaries naturally produce strong double-peaked emission lines \citep{1969AcA....19..155S}, which can superimpose on the P-Cyg profiles. Thus, the blue-shifted absorption of the P-Cyg profile is typically the only wind feature that remains unaltered.
In this study, we consider that a line displays a significant P-Cyg profile when the blue-shifted absorption reaches an intensity of 2\% below the continuum level. H$\alpha$, the most prominent line in our spectra, exhibited evident and variable P-Cyg profiles in epochs \#3 and \#13, and a shallow profile (detected at the limit of our P-Cyg definition) in epochs \#2 and \#16. Figs \ref{sample} and \ref{vlt} show examples of the Balmer emission lines exhibiting P-Cyg profiles. The long ($\sim$6 hours) database obtained on the night of June 3 (epoch \#13) allowed us to study in detail the variability of these features in short time-scales. Fig. \ref{vlt} shows the evolution of H$\alpha$, H$\beta$ and \ion{He}{ii} ($\lambda$4686) emission lines during this epoch. The blue-shifted absorption in H$\alpha$ evolves from a $\sim$2 per cent below continuum level at the beginning of the observing window (\#13-a and \#13-b), to a $\sim$4 per cent in epoch \#13-g (the most prominent P-Cyg profile of the night). The last spectra, epochs \#13-i and \#13-j, show that the line profiles recover a similar shape to those seen at the beginning of the observing window. A visual inspection of these profiles reveals a wind with terminal velocities up to $\sim$2000 km s$^{-1}$, defined as the blue edge of the P-Cyg absorption.

A blue absorption component in H$\beta$ is observed at epochs \#13-g, \#13-h and \#13-i. We suggest that this feature also corresponds to a P-Cyg profile, based on the shape of H$\alpha$ in the same epochs (where the P-Cyg profile is more conspicuous) and the absence of such feature in H$\beta$ in the remaining spectra of the same night. Broad emission line wings and asymmetries are also indicative of the presence of accretion disc winds \citep[e.g.][]{2018MNRAS.481.2646M}. In order to consistently analyse H$\alpha$ emission line wings, we computed the diagnostic diagram developed in \citet{2018MNRAS.481.2646M} and refined in \citet{2019ApJ...879L...4M}. We fitted a Gaussian profile to H$\alpha$, masking the line wings, and then subtracted the fit from the spectrum to estimate the equivalent width of the residuals in both sides of the line. We found that all the data points lie below the 5$\sigma$ confidence level, indicating that these features are either not present or too weak to be detected in our data. For instance, MAXI~J1820$+$070 showed very significant emission line residuals ($\sim$0.45~\AA; \citealt{2019ApJ...879L...4M}) that would sit just above our 3$\sigma$ confidence level. This is due to the lower S/N of our data as compared to that achieved for MAXI~J1820$+$070, whose outburst was significantly brighter.

We observe line asymmetries such as  flat-top profiles (H$\alpha$) in epochs \#3 (see Fig. \ref{sample}) and \#5. These profiles are typically observed in novae and Wolf-Rayet stars and are commonly associated with outflows \citep[e.g.][]{1931MNRAS..91..966B,1954ApJ...119..508V,2010Ap&SS.327..207W}. In addition, we note that the red and blue peaks of the emission lines are highly variable, particularly in epoch \#13 (Fig. \ref{vlt}), with the three lines following the same trend. We observe that the blue peak typically appears to be absorbed (or entirely absent). This is particularly remarkable for the case of \ion{He}{ii} ($\lambda$4686), given that this line has only shown wind signatures in one XRT (Swift J1357.2-0933; \citealt{2019MNRAS.489L..47C,2019MNRAS.489.3420J}; see below). Moreover, this transition exhibits a blue-shifted absorption trough at \mbox{$\sim$-1500~km~s$^{-1}$} (particularly conspicuous in epochs \#13-f to \#13-j) that could be due to the same wind component seen in H$\alpha$ and H$\beta$ as a P-Cyg profile. However, the overlap with the Bowen blend prevents us from extracting more solid conclusions on this matter. 

\subsection{Red-shifted absorptions and inverted P-Cyg profiles}
\label{inflow}
Transient and sharp red-shifted absorptions are observed in epoch \#13 (see arrows in Fig. \ref{vlt}) in both H$\alpha$ and H$\beta$. In analogy to the blue-shifted absorptions of the P-Cyg profiles, these red-shifted features could be identified as inverted P-Cyg profiles associated with inflows of material. For simplicity, from now on we will refer to them as inflow features. They were only present in \#13-c and \#13-d (i.e. for $\sim$105 min). A comparison of the red edge of the emission lines between epochs with and without the inflow feature, indicates that the velocities of these features are higher than the emission line red edges, implying an absorption of the continuum light rather than the disc emission. A Gaussian fit of the inflow features yields a central velocity of $\sim$1000 km s$^{-1}$ and FWHM values of $\sim$200 km s$^{-1}$. Considering the red edge of the feature (measured at the 3$\sigma$ level) as the inflow velocity, we obtain $\sim$1300 km s$^{-1}$. Similar inflow features were also present in epoch \#4 and, less clearly, in epoch \#9 (Fig. \ref{sample}).

In the case of H$\beta$, the inflow feature is superimposed on a static and much broader red-shifted absorption, observed at all epochs and extending up to $\sim$3000 km s$^{-1}$ (Figs \ref{sample} and \ref{vlt}). This is a relatively common feature in XRTs in outburst \citep[e.g.][]{1995ApJ...441..786C,2003MNRAS.342..105B,2007MNRAS.380.1182B} and could be due to a DIB at $\sim$4882~\AA~ \citep{2012ApJ...746L..23K, 1994A&AS..106...39J}. This H$\beta$ feature is most likely not related to the broad absorptions commonly seen in He and Balmer transitions (typically in more than one simultaneously) during some phases of the outburst. These are centred at zero velocity and proposed to be formed when the accretion disc becomes optically thick in these transitions \citep[e.g.][]{2000ApJ...539..445S, 2001ApJ...553..307D}. Nevertheless, it is clear that the transient, narrow and red-shifted absorption present in \#13-c and \#13-d is different from any of the above.   

In Figs \ref{lc} and \ref{hid} we highlight the epochs where inflows and outflows (i.e. inverted P-Cyg and P-Cyg profiles) were detected. Although the remaining epochs lack obvious wind features, we cannot rule out their presence due to the limited quality of the data in some cases. As an example, epoch \#12 corresponds to a SALT observation taken simultaneously with epoch \#13-b (VLT). However, wind signatures are not evident in epoch \#12 due to the decreased S/N ratio. Fig. \ref{hid} shows that both the inflow and outflow detections are concentrated in the top right area of the hardness-intensity diagram, when the source was in the bright hard state.

\section{Discussion}
\label{discussion}
We analysed optical spectra of the BH candidate GRS~1716-249 across the decay of its 2016--2017 outburst. We detected disc wind signatures, such as P-Cyg profiles (or significant absorption of the blue part of the line) and flat-top profiles, in at least 3 (possibly 5) of the 18 observing epochs. We note that the presence/absence of the wind features do not necessarily imply that the wind has a transient nature, i.e. the wind could be steady but only visible under certain conditions. Interestingly, the wind features were only observed in the Balmer and \ion{He}{ii} lines, while the \ion{He}{i} lines, which are typically good tracers of accretion disc winds, were almost absent. We also detected unusual inverted P-Cyg profiles in 3 epochs, implying the likely additional presence of an inflow. 

Cold (optical-infrared) accretion disc winds have been detected in a number of sources: V404 Cyg \citep{1991MNRAS.250..712C,2016Natur.534...75M,2017MNRAS.465L.124M}, V4641 Sgr \citep{2003MNRAS.343..169C,2018MNRAS.479.3987M}, MAXI J1820$+$070 \citep{2019ApJ...879L...4M, 2020arXiv200707257S}, Swift J1357.2-0933 \citep{2019MNRAS.489L..47C,2019MNRAS.489.3420J}, and Swift J1858.6-0814 \citep{2020ApJ...893L..19M}. GRS~1716-249 is therefore the sixth XRT discovered to display unambiguous signatures of low-ionisation accretion disc winds. The terminal velocity of the outflow ($\sim$2000~km~s$^{-1}$) is in agreement with those observed in the other systems. The observed absorption in the blue part of the \ion{He}{ii} ($\lambda$4686) emission line (only seen before in Swift~J1357.2-0933), simultaneous with P-Cyg profiles in the Balmer lines, points to a hot and dense wind \citep[see][for a discussion on this matter]{2019MNRAS.489.3420J,2019MNRAS.489L..47C}. There is strong evidence supporting a high orbital inclination for Swift~J1357.2-0933 \citep[see also][]{2013Sci...339.1048C}. However, this parameter is unknown for GRS~1716-249, for which the presence of high inclination features have not yet been reported despite the 2016--2017 outburst being particularly well covered.
GRS~1716-249 remained in the hard state during the entire outburst \citep{2019MNRAS.482.1587B}, as was the case for the 1993 event \citep{1998A&A...331..557R}. Its evolution in the hardness-intensity diagram reveals that the optical wind signatures are conspicuous in the bright hard state, which is consistent with that observed in MAXI J1820$+$070 \citep[][]{2019ApJ...879L...4M}.

\subsection{The origin of the inflow features}
\label{disc_inf}
We observed narrow red-shifted absorptions (so-called inflow features) in H$\alpha$ and H$\beta$ simultaneously. This implies that these features are clearly independent from the static and broad red-shifted absorptions present in H$\beta$ that we discuss in Section \ref{inflow}. One possible scenario for the red-shifted narrow absorptions is an inflow of material like a failed wind, i.e. outflowing material that falls to the disc again because it does not reach the escape velocity. Considering that our estimated outflow terminal velocity ($\sim$2000~km~s$^{-1}$) is equivalent to the Keplerian velocity (which is 1/$\sqrt{2}$ times the escape velocity) at the launching radius R$_\mathrm{l}$, and taking a typical BH mass of 8 M$_{\odot}$, we obtain R$_\mathrm{l} \simeq 2.25 \times$10$^{4}$ R$_\mathrm{g}$, where R$_\mathrm{g}$ is the gravitational radius. Assuming that the infalling gas has a free-fall velocity, we can estimate the line formation radius in an analogous way. The inflow velocity ($\sim$1300~km~s$^{-1}$) yields an inflow radius R$_\mathrm{inf} \simeq 5.32 \times$10$^{4}$ R$_\mathrm{g}$. 
In order to compare these radii with the size of the accretion disc, we can roughly estimate the outer disc radius. Assuming a typical mass ratio of $q<$~0.1 \citep[see][]{2016ApJ...822...99C,2016A&A...587A..61C} and an orbital period of 5~hours~--~1~day, the tidal truncation radius gives an upper limit to the disc radius of $\gtrsim (1-3) \times$10$^{5}$~R$_\mathrm{g}$, while the circularisation radius gives a lower limit of $\gtrsim (0.6-1.6) \times$10$^{5}$~R$_\mathrm{g}$ \citep[for further details on these calculations see][]{2020ApJ...893L..37T}. These approximations locate both the inflow and outflow characteristic radii within the limits of the accretion disc.

Instead of being caused by a failed wind, the inflow features could be due to a persistent accretion component only visible at certain orbital phases, such as the gas stream from the donor star. This scenario would be possible if the system is seen close to edge-on. However, although the wind properties are compatible with a high orbital inclination, neither eclipses nor dips have been reported to date. In addition, we would expect to see changes in the velocity of the feature due to projection effects, but the inflow velocity seems to be rather static in epoch \#13-c/d. Moreover, the relative strength of the blue and red peaks of the emission lines show sometimes an orbital dependence. While the inflow feature at epoch \#13-c/d coincides with a strong red peak in both H$\alpha$ and H$\beta$ (Fig. \ref{vlt}), that of epoch \#9 concurs with a slightly stronger blue peak (Fig. \ref{sample}), suggesting that the events took place at different orbital phases. In addition, the higher excitation lines [i.e. the Bowen blend and \ion{He}{ii} ($\lambda$4686)] are good tracers of the amount of irradiation received by the system and the visibility of the more irradiated regions. Fig. \ref{vlt} shows that these lines are stronger (and tend to display sharper components) during the first part of epoch \#13 (including \#13-c and -d). One could speculate that this is due to a higher visibility of the irradiated donor or the bulge. The latter would point towards orbital phases $\sim$0.6--0.8. In a high inclination scenario, the bulge could obscure part of the receding portion of the disc at these phases. However, as noticed in Section \ref{inflow}, the red-shifted absorptions are beyond the edge of the disc emission components. In any case, we note that the orbital period is uncertain for this source, which prevents a more thoughtful consideration of this scenario.

We note that gravitational redshift could in principle be considered as a possible scenario to explain the red-shifted narrow absorption. By using Eq. 5 from \citet{2013MNRAS.434..222H}, that gives the gravitational redshift for a transition around a Schwarzschild BH, we infer a line formation radius of $\sim$200~R$_\mathrm{g}$ (for $\sim$1300 km s$^{-1}$). This calculation therefore argues against a gravitational redshift explanation for the observed inflows, as gas at such a small radius is expected to have a much higher temperature and thus it is not traced by the optical emission.

Although inflow features in XRTs are very unusual, they have tentatively been detected before. \citet{2014ApJ...788...53M} reported on Fe absorption features in the X-ray spectra of MAXI~J1305-704 and suggested that they could be related to a failed X-ray wind. Despite their scarcity among XRTs, inflow features are common in active galactic nuclei (AGN). Red-shifted absorptions in H and He optical lines have been observed in quasars \citep{2019Natur.573...83Z}. These features have also been witnessed in other spectral bands, such as the infrared \citep[e.g. in the OH 119 $\mu$m molecular doublet of Circinus,][]{2016ApJ...826..111S}, UV \citep[e.g. in broad absorption line quasars,][]{2013MNRAS.434..222H}, and X-rays \citep[e.g. in CID-42,][]{2010ApJ...717..209C,2013ApJ...778...62L}. A number of different scenarios to explain how inflow features are produced in AGN have been proposed, including rotationally dominated outflows, gravitational redshift, binary quasars, relativistic Doppler shift and infall of gas close to the BH at velocities near the free-fall speed. Moreover, failed winds are predicted by simulations of radiation-driven disc outflows in luminous AGN \citep[e.g.][]{2019A&A...630A..94G,2020arXiv200207564G}.

\subsection{On the wind launching mechanism}
Different wind-launching mechanisms have been discussed in the literature, including radiation-driven winds (Thompson scattering), thermal outflows \citep[Compton-heated wind;][]{1983ApJ...271...70B} and magnetic winds \citep[i.e. launched by magnetic pressure; e.g.][]{2006Natur.441..953M}. 

A wind driven by radiation pressure is only possible close to the Eddington limit \citep[e.g.][]{2007MNRAS.377.1187P}, which is not often the case for X-ray binaries (but see \citealt{2002A&A...391.1013R} and \citealt{2019MNRAS.488.1356C} for possible observational evidences in V4641 Sgr and V404 Cyg, respectively). However, some (resonant UV) transitions are able to trigger winds at luminosities well below the Eddington limit; these are line-driven winds. It has been shown that the line-driven mechanism is not expected to be at work in X-ray binaries because the gas surrounding the compact object would be entirely ionized by the strong X-radiation \citep{2002ApJ...565..455P}. However, this scenario becomes plausible if there is shielding material preventing the gas from being overionized \citep[e.g.][]{2017MNRAS.468..981M} and, as a matter of fact, we observed wind features likely arising in the (low-ionized) outer accretion disc. In this framework, if the inflow features are caused by a failed wind, this could result from inner disc material that cannot overcome the gravitational pull from the central BH and falls back \citep[][see also \citealt{2019A&A...630A..94G}]{2020arXiv200207564G}. These authors even suggested that failed winds might act themselves as a shielding structure. However, the inflow velocity derived in this work ($\sim$1300 km s$^{-1}$) does not seem to be consistent with the inner disc. 

Compton heated winds are produced when disc material reaches a thermal velocity higher than the escape velocity. If part of this gas does not reach the velocity to become fully unbound, it could fall back with the free-fall velocity, producing an inflow. Therefore, the inflow velocities are smaller than the outflow velocities, which is what we observe in GRS~1716-249. However, the characteristics of the  observed inflow features, such as their transient appearance, their short duration (e.g. as compared to the outflow features) and narrowness, suggest that the inflow is dense and stratified, properties that may not fit with the very simplistic scenario described above \citep[for simulations on the physical properties of thermal winds see][]{2017ApJ...836...42H,2019MNRAS.484.4635H,2020MNRAS.492.5271H,2018MNRAS.473..838D}.

Finally, magnetic processes have been widely discussed in the literature as a possible mechanism to launch winds in BH transients \citep[e.g.][]{2006Natur.441..953M,2018MNRAS.481.2628W}. Although this scenario is not well understood yet, a magnetic process might account for an organized inflow like that detected here in GRS~1716-249.

\section{Conclusions}
\label{summary}
We have presented optical spectroscopy of GRS~1716-249 during the decay of its 2016--2017 outburst. We observe P-Cyg profiles and other outflow signatures indicating the presence of an accretion disc wind (at $\sim$2000 km s$^{-1}$), which is likely hot and dense. In addition, we detect inflow features (narrow red-shifted absorptions at $\sim$1300 km s$^{-1}$) that might be caused by a failed wind or a persistent accretion structure such as the gas stream from the donor star, albeit the latter scenario requires a high orbital inclination. Given that typical observing campaigns are usually limited to a few spectra per night over a limited number of epochs, the short duration (or limited visibility) of the inflow features implies that they are difficult to be observed. Future monitoring programs of XRTs making use of relatively long observing windows ($\sim$hours), should allow us to investigate whether GRS~1716-249's behaviour is unique, or if inflows are also a common property of XRTs, alongside the frequent wind-type outflows.

\section*{Acknowledgements}
We acknowledge support from the Spanish \textit{Ministerio de Ciencia e Innovaci\'on} under grant AYA2017-83216-P. TMD and MAPT acknowledge support via Ram\'on y Cajal Fellowships RYC-2015-18148 and RYC-2015-17854. DAHB acknowledges support from the National Research Foundation of South Africa. JMC-S acknowledges financial support to CONICYT through the FONDECYT project No. 3140310. DMS acknowledges support from the ERC under the European Union's Horizon 2020 research and innovation programme (grant agreement no. 715051; Spiders). GP is supported by the H2020 ERC Consolidator Grant programme (grant agreement Nr. 865637; Hot Milk). \textsc{molly} software developed by Tom Marsh is gratefully acknowledged. We made use of observations collected at SALT under program 2016-2-LSP-001, GTC under program GTC34-17A and ESO under program 299.D-5013(A). Polish participation in SALT is funded by grant no. MNiSW DIR/WK/2016/07. We acknowledge \textit{Swift}/BAT transient monitor results provided by the \textit{Swift}/BAT team and \textit{MAXI} data provided by the RIKEN, JAXA, and MAXI teams.

\section*{Data Availability}
The data underlying this article are publicly available in: \textit{Swift} \url{https://swift.gsfc.nasa.gov/results/transients/}; \textit{MAXI} \url{http://maxi.riken.jp/top/slist.html}; VLT [program 299.D-5013(A)] \url{http://archive.eso.org/cms.html}; GTC (program GTC34-17A) \url{https://gtc.sdc.cab.inta-csic.es/gtc/}. SALT data (program 2016-2-LSP-001) can be requested through \url{https://astronomers.salt.ac.za/data/download-data-and-calibrations/}.




\bibliographystyle{mnras}
\bibliography{grs1716} 








\bsp	
\label{lastpage}
\end{document}